\documentclass{article}
\usepackage{spconf,amsmath,graphicx}
\usepackage{epsfig,url,comment}
\usepackage{enumitem, multirow}

\title{BUT OpenSAT 2019 speech recognition system}
\name{Martin Karafi\'{a}t, Murali Karthick Baskar, Igor Sz\"{o}ke, Hari Krishna Vydana, Karel Vesel\'{y}, Jan ``Honza'' \v{C}ernock\'{y}
  \thanks{The work  was supported by Czech Ministry of Education,
    Youth and Sports from the National Programme of Sustainability
    (NPU II) project "IT4Innovations excellence in science - LQ1602"
    and by the Office of the Director of National Intelligence (ODNI),
    Intelligence Advanced Research Projects Activity (IARPA) MATERIAL
    program, via Air Force Research Laboratory (AFRL) contract \#
    FA8650-17-C-9118. The views and conclusions contained herein are
    those of the authors and should not be interpreted as necessarily
    representing the official policies, either expressed or implied,
   of ODNI, IARPA, AFRL or the U.S. Government.}
   }
\address{Brno University of Technology Speech@FIT and 
  IT4I Center of Excellence \\
  Brno, Czechia}
\begin{document}
\ninept
\maketitle
\begin{abstract}
The paper describes the BUT Automatic Speech Recognition (ASR) systems submitted for OpenSAT evaluations under two domain categories such as low resourced languages and public safety communications. The first was challenging due to lack of training data, therefore various architectures and multilingual approaches were employed. The combination led to superior performance. The second domain was challenging due to recording in extreme
conditions such as specific channel, speaker under stress and high levels of noise. Data augmentation process was inevitable to get reasonably good performance.
\end{abstract}

\section{Introduction}
\label{sec:intro}

The paper focuses on our work on building the ASR system for the
Speech Analytic Technologies (OpenSAT) evaluation running  in summer
2019 an organized by National Institute of Standards and Technology (NIST). It focused on three tasks:
 Automatic Speech Recognition (ASR),  Speech Activity Detection (SAD) and Keyword Search (KWS)
and on three domains:
Low resourced languages (Babel), speech from video - (VAST, with  SAD task only) and Public Safety Communications (PSC).  

Our team worked on the ASR task for the following domains: Low resourced languages and public safety communications.
While the  first one is challenging by limited resources, the second one is addressing English but contains a lot of speech under stress 
and recorded through specific channel. To be successful in both challenges, several  acoustic
models were built (section~\ref{sec:AM}). 
It is known that ASR systems tend to produce system-dependent errors,
therefore complementarity and system fusion is crucial for good performance. 
Consequently, we focused on building more systems which vary in many
parts to increase their complementarity: different acoustic units,
training data (monolingual/multilingual, clean/noise), architectures
(with/without convolution layers), acoustic modeling approaches
(hybrid/end-to-end) and bandwidth (sampling frequency 8/16kHz).

The detailed system
descriptions are given in sections ~\ref{sec:babel} and~\ref{sec:psc} and
results are presented in sections~\ref{sec:babelresults} and~\ref{sec:PSCresults}.

\subsection{Low Resource Languages}
Pashto language was selected for this challenge from 
IARPA Babel program (running from 2012 to 2015)~\cite{harper:ASRU:2013}, therefore this challenge
is later noted as Babel. 
In Babel, data from various low-resource
languages were collected, which allowed us to focus on multilingual 
experiments and acoustic
modeling. It is very natural for humans to borrow information from other
sources when trying to learn a new language. Humans
share the same vocal tract architecture and phonetic systems overlap across many
languages, therefore automatic
systems should be able to have components built and trained on
various data sources.

During the Babel project, where we worked as part of ``Babelon'' team (led by BBN), we verified that multilingual pre-training for feature extraction  is an important
technique especially if not enough training data is available~\cite{grezl:SLTU:2014:BNmultNN}. 
Recently, we also extended multilingual Deep Neural Network (DNN) acoustics models
~\cite{ghoshal:ICASSP:2013:MultilingDNN} to 
novel NN architectures~\cite{karafiat:IS:2017:MultBLSTM} and
presented significant gains with adding more languages into acoustic
model pre-training. 

Consequently, we focused on system fusion of monolingual and multilingual models as we expected complementarity
from various training data. Moreover, we built the systems with various architectures and modeling units, see~\ref{sec:babelAM} for details.

\subsection{Public Safety Communications}
On contrary to the pilot OpenSAT evaluations conducted in
2017~\cite{karafiat:IS:2018:OpenSAT}, where the target data was real
fireman-dispatcher communication  from the Charleston Sofa Super Store
Fire in 2007, for OpenSAT 2019, NIST prepared simulated public safety communications collection ("SAFE-T") specifically designed for speech analytic systems. The data is
intended to simulate a combination of characteristics found in public
safety communications: background noises, channel transmission noises and
speaking characteristics such as stress or sense of urgency.

In order to utilize additional corpora, like clean telephone data, we  
also experimented with simulated channel (see section~\ref{sec:channel} attempting to close  the  gap
between target and clean training data. Finally, we focused on system fusion of various models trained only on "SAFE-T" data as well as models transferred from clean to target domain.

\section{Acoustic models and techniques}
\label{sec:AM}

\subsection{Hybrid Acoustics models}
Various classical hybrid DNN-HMM (Hidden Markov Model) speech recognizers were trained in Kaldi
toolkit~\cite{povey:ASRU:2011:Kaldi}. We decided to select Factorized
Time Delay NN ({\bf TDNN-F}) based
architectures~\cite{povey:IS:2018:TDNN-F} as we were consistently
obtaining superior performance to recurrent NN architectures used in last OpenSAT evaluations~\cite{karafiat:IS:2018:OpenSAT}:
\begin{itemize}
\item {\bf TDNN-F}: 15 layers with 1536 neurons, with bottle-neck factorization to 60 dimension and with stride 3.
\item {\bf CNN-TDNN-F}: Similar to {\bf TDNN-F} above, only the first 6 layers are replaced by convolutional ones.
\end{itemize}

The feature extraction was based on high-resolution MFCC features (40dim)
concatenated with 100 dimensional online
ivectors~\cite{peddinti:IS:2015:ivec-aspire-online}.
Next, the features were subsampled by factor 3 and NNs were trained with
Lattice Free MMI (LF-MMI) objective and bi-phone targets as suggested
in~\cite{Povey:IS:2019:LF-MMI}. 
Finally, the NN are further trained with sequence Minimum Bayes Risk sMBR
objective~\cite{vesely:IS:2013:DNN_sMBR}.  

In order to increase the variety of our systems, we experimented with
different units as acoustic model training targets:
In addition to the bi-phone NN targets, which were shown to be
on-par with the classical tri-phone ones~\cite{Povey:IS:2019:LF-MMI}, we experimented with single-state context independent phonemes 
as the NN model training targets.

\subsection{Multilingual hybrid models}
\label{sec:multAM}

In previous
evaluations~\cite{karafiat:IS:2017:MultBLSTM,karafiat:IS:2018:OpenSAT}
we were using block-softmax objective function~\cite{vesely:slt:2012}
which is not yet implemented with LF-MMI training in Kaldi. Therefore,
we chose a much simpler approach: The training language data (audio lists,
transcriptions, pronunciation dictionaries...)  are simply stacked together with
added language indicator for phoneme names  as suggested
in~\cite{Keith:ICASSP:2018:MultilinualTDNN}. This approach is ensuring
non mixing of acoustic units across the languages. Then, the training
of multilingual NN is same as monolingual one.  

Procedure of porting the multilingual NN into the new language can be
described in the following steps:
(1) the  final multilingual layer (stacked all language-dependent phoneme units)
 are stripped and replaced with a new layer specific to
target-language with random initialization. (2) Learning rate of the rest of NN is scaled
by 0.5. (3) The whole NN is retrained by two epochs on the target language.

\subsection{Sequence-to-sequence model}
With increasing interest in Sequence-to-sequence (seq2seq)
models~\cite{chorowski2015attention,watanabe2017hybrid}, we decide to
experiment with them in order to obtain complementary approach to classical hybrid DNN-HMM.  

Joint CTC/Attention \cite{watanabe2017attctc} based sequence-to-sequence ASR system is used in this
work. The pipeline contains an input layer with a 2D-convolution layer followed by 12 multi-head self-attention~\cite{vaswani2017attention} encoder layers. The 2D convolution layer sub-samples the input by factor of 4. The encoded input is fed to 12 multi-head restricted self-attention decoder layers.  The restricted self-attention~\cite{yang2019convolutional} is enforced using a 1D convolution component to focus only the neighboring region. 

\subsection{Voice Activity detection}
Voice Activity detection (VAD) system was based on feed-forward Neural Network with 2 outputs.
Standard mel-filter bank features (15 coefficients) with F0 estimates (3 coefficients)~\cite{povey:ICASSP:2014:F0} were taken as the input after cepstral mean and variance normalization.
Based on the training data we trained two VAD systems 
\begin{itemize}
    \item VAD-Babel: trained on publicly available Babel data (see section~\ref{sec:babeldata}).
    \item VAD-Safe-T: trained on downsampled (8kHz) PSC corpus (see section~\ref{sec:pscdata}).
\end{itemize}

\subsection{Score combination}
ROVER tool from SCTK 
package\footnote{
\url{http://www1.icsi.berkeley.edu/Speech/docs/sctk-1.2/sctk.htm}
}
is used to perform score combination as a fusion of system outputs. To
improve Normalized Cross Entropy (NCE) scores, logistic regression model is 
built to perform calibration. System word confidence and the language
model score is provided as input to the training of the logistic regression model.

\subsection{PSC data enhancement}
\label{sec:channel}
The following methods of data enhancement were applied on the clean training data:
\begin{enumerate}
  \item Adding reverberation (all training data).
  \item Adding environmental noise (all training data).
  \item Passing through walkie-talkie ``channel'' (1/2 of the training data).
  \item Passing through codecs (all training data).
\end{enumerate}

The audio data were processed on a segment chunks level. We used BABEL-VAD
for speech/non-speech segmentation. Next, we made segment chunks of length $5$ seconds (or more) by grouping particular segments in sequence. The chunk boundary is an event to change (randomly) the augmentation parameters -- impulse response for reverberation,  additive environmental noise, and SNR level. The motivation is to enlarge the speaker and environment variability.

The other set of augmentation steps is performed on level of the whole recordings.

\subsubsection{Adding speech reverberation}
We selected $2684$ impulse responses from ACE ($452$), AIR $1.4$
($52$), REVERB ($104$), RWCP ($1921$), ReverbDB ($155$) datasets
heaving RT$60$ lower than $0.5$ seconds. 

\subsubsection{Adding environmental noise}
We used various noise sources:
\begin{itemize}
    \item Sounds downloaded from \texttt{freesound.org} ($463$ minutes in total) consisting of: 1) $37$ minutes of engine noises, 2) $330$ minutes of airplane cabin noises, 3) $96$ minutes of HAM radio static noises.
    \item Environmental empty room noises from ReverbDB dataset ($1054$ minutes in total)
\end{itemize}

We mixed the data with $4$ x freesound ($1852$ minutes) + $1$ x
reverbDB ($1054$ minutes).

Next, each chunk of segments was corrupted by a single, randomly selected noise picked from pool of $1452$ samples of freesound ($463$ minutes $\times 4 = 1852$ minutes) and $1054$ samples of ReverbDB ($1054$ minutes $\times$ $1 = 1054$ minutes). 
The starting position of the noise segment was randomly selected. 
The target SNR was chosen randomly from the interval $1-8$dB (low) and $9-15$dB (mid).

\subsubsection{Walkie-talkie ``channel''}
We tried to simulate the HW (radio station) and ambiance in the following way:
\begin{enumerate}
  \item Normalization of the training audio to $0$ dB gain.
  \item Increasing gain from $0$dB to $20$dB to introduce clippings.
  \item Application of randomly selected high-pass filter with cut-of
    frequency of $300,600,1000,$ or $1500$Hz.
  \item Application of the phaser effect with ``sox'' tool to simulate phase distortions.
\end{enumerate}

\subsubsection{Codecs}
The fire departments in the U.S. use the proprietary AMBE codec in
their Digital Mobile Radio (DMR). The source code of this codec is not
available. As a replacement, we used the EU version of the TETRA codec
to simulate the effects of signal coding. We assume that the TETRA
codec has characteristics similar to the AMBE one. We added also AMR, AMR-WB, G.711, G.726, G.729, GSM-FR, GSM-EFR, MP3 codecs with various bitrates (lower values) and up to $6\%$ frame drop rates if possible.

\section{BABEL system description}
\label{sec:babel}

\subsection{Data}
\label{sec:babeldata}
Multilingual acoustic models were pre-trained on 20 Babel
Languages (all languages available from Linguistic Data Consorcium - LDC:  
Cantonese, Pashto, Turkish, Tagalog, Vietnamese,
Assamese, Bengali, Hait. Creole, Lao, Swahili, Georgian, Tamil, Zulu, Kurdish Kurmanji, Tok Pisin, Cebuano, Kazakh, Telugu, Lithuanian, Guarani.
The monolingual system was trained only on provided Pashto  data-set (cca 92~hours).  

\begin{table}[ht]
\begin{tabular}{|l|l|l|}
\hline
Cantonese & LDC2016S02 & IARPA-babel101-v0.4c \\
Assamese  & LDC2016S06 & IARPA-babel102b-v0.5a\\
Bengali   & LDC2016S08 & IARPA-babel103b-v0.4b\\
Pashto    & LDC2016S09 & IARPA-babel104b-v0.4Yb\\
Turkish   & LDC2016S10 & IARPA-babel105-v0.6 \\
Georgian & LDC2016S12  & IARPA-babel404b-v1.0a \\
Tagalog   & LDC2016S13 & IARPA-babel106-v0.2g \\
Vietnamese& LDC2017S01 & IARPA-babel107b-v0.7 \\
Haitian Creole         & LDC2017S03 & IARPA-babel201b-v0.2b\\
Swahili  & LDC2017S05  & IARPA-babel202b-v1.0d \\
Lao      & LDC2017S08  & IARPA-babel203b-v3.1a \\
Tamil    & LDC2017S13  & IARPA-babel204b-v1.1b \\
Zulu     & LDC2017S19  & IARPA-babel206b-v0.1e \\
Kurmanji & LDC2017S22  & IARPA-babel205b-v1.0a \\
Tok Pisin & LDC2018S02 & IARPA-babel207b-v1.0e \\
Cebuano  & LDC2018S07  & IARPA-babel301b-v2.0b \\
Kazakh   & LDC2018S13  & IARPA-babel302b-v1.0a \\
Telugu   & LDC2018S16  & IARPA-babel303b-v1.0a \\
Lithuanian & LDC2019S03 &  IARPA-babel303b-v1.0a \\
Guarani  & LDC2019S08  & IARPA-babel305b-v1.0c \\
\hline
\end{tabular}
  \caption{Used Languages for multilingual pre-training.}
  \label{tab:data}
\end{table}

The test and evaluation data were processed by {\bf VAD-BABEL} and
decoded with standard 3-gram back-off ARPA language model trained on acoustic model training data transcriptions and on WEB data downloaded during Babel program~\cite{zhang:IS2015:webdata}.

\subsection{Acoustic models}
\label{sec:babelAM}

The following models were trained for final fusion:
\begin{itemize}
\item Baseline\_CNN-TDNN-F: CNN-TDNN-F based system trained purely on Pashto data
\item Baseline\_CNN-TDNN-F\_phnout: CNN-TDNN-F using \\ phoneme target units
\item Multilingual\_CNN-TDNN-F: CNN-TDNN-F system pre-trained on 20
  languages and transferred to the target language.
\item Multilingual\_TDNN-F: TDNN-F system pre-trained on 20
  languages and transferred to the target language.
\end{itemize}

\subsection{Results}
\label{sec:babelresults}

\begin{table}[tb]
\caption{WER on development data.}
\label{tab:BABELwer}
\begin{center}
\begin{tabular}{lc}
\hline
System & WER[\%] \\
\hline
Baseline  & 39.7 \\
Baseline\_phnout & 40.3 \\
Multilingual\_CNN-TDNN-F & 39.1 \\
Multilingual\_TDNN-F & 39.0 \\
\hline
Fusion  & {\bf 37.4} \\
\hline
\end{tabular}
\end{center}
\end{table}

Table~\ref{tab:BABELwer} presents performances on the development (dev) set. We
observed tiny (0.6\%) gain from using multilingual acoustic model over the monolingual one. It is probably
due to the sufficient amount of training data (92~hours). Phoneme
based acoustic model shows only 0.6\% degradation over bi-phone
baseline so we deem it suitable system for fusion. Moreover, it shows strong modeling power of NN based system trained with LF-MMI, compared to old Gaussian Mixture models where at least (bi/tri)-phonemes were needed to get reasonable performance. The final fusion shows significant 2.3\% gain over monolingual baseline. 
This fusion also performed well on the evaluation data, see `Babel' part of figure~\ref{fig:eval} to compare with other participants in the OpenSAT Evaluation.

\section{PSC system description}
\label{sec:psc}
\subsection{Data}
\label{sec:pscdata}
Various English data-sets were used for acoustic model training:
\begin{itemize}
    \item {\it FULL-NB-CLEAN} (2240 hours): English Fisher1+2, 
Switchboard 1 Release 2, Call Home English, AMI and ICSI-meetings. The
wide-band corpora (AMI and ICSI) were downsampled to 8 kHz.
    \item {\it FULL-NB-NOISE} (2240 hours): {\it FULL-NB-CLEAN} set further processed by channel enhancement (see section~\ref{sec:channel}).
    \item {\it MEETING-WB} (174 hours): Only wide-band (WB) parts of {\it
      FULL-NB-CLEAN} (AMI, ICSI) with original sampling rate 16 kHz.
    \item {\it SAFE-T-NB} (20.6 hours):
      Target SAFE-T corpora downsampled to 8 kHz.
    \item {\it SAFE-T-WB} (20.6 hours):
      Target SAFE-T corpora with the original 16 kHz sampling rate.
\end{itemize}

For the testing, we used 3-gram Language Model based on SAFE-T transcriptions. In
addition, we trained RNN-LM (with 5 layers: TDNN, LSTM, TDNN, LSTM, TDNN)
on the same data. It was used for final lattice re-scoring.
Note, we experimented with RNN-LM also for Babel domain without any gain. Therefore, this technique is used only in PSC domain.

\subsection{Acoustics models}
\label{sec:pscAM}

\noindent {\bf Various pre-training methods:}
We experimented with two techniques for porting of models trained on
additional corpora ({\it FULL-NB-NOISE}, {\it FULL-NB-CLEAN}) data into target {\it SAFE-T-NB} one: 
\begin{enumerate}
    \item {\bf model-transfer}: Last layer is rebuilt on target data
      in the same way as in language-transfer learning used in porting
      of multilingual NN (see section~\ref{sec:multAM}), just the
      training/target language is same.
    \item {\bf fine-tuning}: further model training (4 epochs) on target data with scaling of learning rate by factor 0.05. 
\end{enumerate}

\begin{table}[tb]
\caption{PSC: comparison of various pre-training methods.}
\vspace{2mm}
\label{tab:psc:am-pretraining}
\begin{center}
\begin{tabular}{lcc}
\hline 
Pre-training NN & Method & WER[\%] \\
\hline
None (SAFE-T-NB only) & - &  17.0 \\ 
 FULL-NB-CLEAN & model-transfer & 15.8  \\
 FULL-NB-NOISE & model-transfer & 15.3  \\
 FULL-NB-NOISE & fine-tuning    & {\bf 14.2}  \\
\hline
\end{tabular}
\end{center}
\end{table}

Table~\ref{tab:psc:am-pretraining} presents comparison of these two
approaches on LF-MMI CNN-TDNN-F models. The development data were
processed with VAD-Babel, no RNN-LM and no sMBR training was used for
simplicity. It shows significant 1.2\% gain from using additional
training data sources. Using noised data gives another 0.5\%
improvement. Simple fine-tuning instead of mode-transfer is taking
advantage from pre-trained last layer (in model transfer the last
layer is rebuilt on target data with random initialization) by giving
additional 1.2\%. 
Therefore, simpler fine-tuning was selected as model-transfer method for later wide-band experiments where {\it MEETING-WB} was utilized for the training 
(see section~\ref{sec:PSCresults}).

Next, additional effect of training with sMBR and RNN-LM lattice rescoring is presented
in table~\ref{tab:psc:sMBR_RNNLM}. It shows  
tiny 0.2\% improvement by sMBR and 0.8\% additional gain by RNN-LM.

\begin{table}[ht]
\caption{PSC: effect sMBR and RNN-LM on fine-tuned FULL-NB-NOISE models.}
\label{tab:psc:sMBR_RNNLM}
\begin{center}
\begin{tabular}{lc}
\hline 
System & WER[\%] \\
\hline
Initial (LF-MMI CNN-TDNN) & 14.2  \\ 
  + sMBR       &   14.0 \\
 ~~ + RNN-LM   & 13.2  \\
\hline
\end{tabular}
\end{center}
\end{table}

\noindent {\bf Sequence-to-sequence training}
Experiments with sequence-to-sequence training were conducted using EspNET~\cite{watanabe2018espnet} toolkit. 83 dimensional log-Mel filter-bank features are used as input and 5002 sentence-piece units are used as output labels. The encoder and decoder contain 2048 dimensional hidden units and 4 attention heads. CTC weight of 0.3 is applied during training and decoding. The model training is regularized with  dropout (0.1) and label smoothing (0.1). Spectral augmentation is applied by time warping and frequency masking. 
The final model is obtained by averaging the best 10 models. The performance of the system is denoted in table~\ref{tab:PSCwer}.  

\subsection{Results}
\label{sec:PSCresults}

The following models were used for final fusion:
\begin{itemize}
    \item {\bf SAFE-T-NB} CNN\_TDNN-F baseline trained in 8kHz sampling sampling rate      
    \item {\bf SAFE-T-WB} CNN\_TDNN-F baseline trained in 16kHz sampling sampling rate      
    \item {\bf FULL-NB-NOISE}: CNN\_TDNN-F model pre-trained on {\it FULL-NB-NOISE} data and
      further trained on {\it SAFE-T-NB} (3 epochs).
    \item {\bf MEETING-WB}: CNN\_TDNN-F model pre-trained on {\it MEETING-WB} data and further
      trained on {\it SAFE-T-WB} (3 epochs).
    \item {\bf Seq2seq-WB}:  Seq2seq model pretrained on 960 hours of Librispeech data and fine-tuned to
    {\it SAFE-T-WB} data.
\end{itemize}
\begin{table}[ht]
\caption{PSC: final systems results.}
\vspace{2mm}
\label{tab:PSCwer}
\begin{center}
\begin{tabular}{lcc}
\hline 
System & \multicolumn{2}{c}{WER[\%]} \\
       & VAD-Babel & VAD-SAFE-T        \\
\hline
 SAFE-T-NB     & 16.4 & 15.3 \\ 
 SAFE-T-WB     & 15.0 & {\em (2)} 14.0 \\
 FULL-NB-NOISE & {\em (1)} {\bf 13.2} & 13.7 \\
 MEETING-WB    & 13.8 & 14.1 \\
Seq2seq-WB            & {\em (3)} 16.5 & -    \\ 
\hline
fusion {\em (1)}+{\em (2)}    & \multicolumn{2}{c}{\multirow{2}{*}{12.4}} \\
(no score calibration) & & \\
fusion {\em (1)}+{\em (2)}             & \multicolumn{2}{c}{12.2} \\
fusion {\em (1)}+{\em (2)}+{\em (3)}   & \multicolumn{2}{c}{11.7} \\ 
fusion all (no score calibration)      & \multicolumn{2}{c}{11.1} \\
fusion all                             & \multicolumn{2}{c}{\bf 11.0} \\
\hline
\end{tabular}
\end{center}
\end{table}

Table~\ref{tab:PSCwer} presents 1.3-1.4\% absolute improvement using
original WB data instead of downsampling to 8kHz. Training with help of noised data (8kHz) is getting additional 0.3-1.8\% gain over wide-spectra training and presents 1.6-3.2\% improvement over 8kHz baseline. Therefore, {\bf FULL-NB-NOISE} based system was our single best system. Unfortunately, we did not have time to create {\bf MEETING-WB-NOISE} data-set, therefore we fine-tuned only clean {\bf MEETING-WB} generating intermediate results. 

\begin{figure}[tbh]
  \begin{center}
    \epsfig{figure=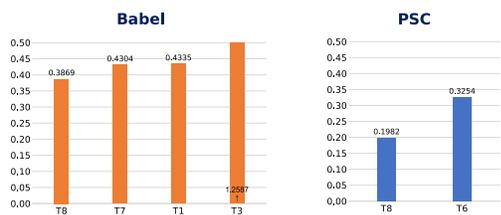,height=0.20\textheight} 
    \vspace{-10mm}
    \caption{Evaluation results in WER ("T8" is BUT system).}
    \label{fig:eval}
  \end{center}
  \vspace{-8mm}
\end{figure}

Our systems in PSC section vary a lot, therefore we run a detailed
analysis with focus on complementarity. The most complementary system
to the best single system, mark as {\em (1)} in table~\ref{tab:PSCwer}, was {\bf SAFE-T-WB} {\em (2)} processed by
SAFE-VAD, probably due to different bandwidth and
segmentation. End-2-End system performs close to hybrid baseline and
helps in system fusion by additional 0.5\% over {\em(1)}+{\em(2)}. Fusion of all systems provides 0.7\% additional gain and shows 2.2\% absolute improvement over the best single system. In addition,
we present small 0.1-0.2\% gain by score calibration. Finally, we
got really good results on the eval data in comparison to the other evaluation participant (see PSC part of figure~\ref{fig:eval}).

\section{Conclusion}

The paper presented our efforts for OpenSAT evaluation 2019. We
participated in two domains: Low Resource Languages and Public Safety
Communications. In the first one,  we confirmed the importance of creating
highly  complementary systems  as well as using multilingual approaches.
In the second one, we described our specific training data enhancements
and investigated complementarity with recent end-to-end based system. In
both domains, our systems had the best performance in the OpenSAT Evaluations.



\bibliographystyle{IEEEbib}
\bibliography{karaf,karthick}

\end{document}